\documentclass{mypnas}

\usepackage[pdftex]{graphicx}
\usepackage[usenames,dvipsnames]{color}
\usepackage{amssymb,amsfonts,amsmath}
\usepackage{booktabs}
\usepackage{dcolumn}
\renewcommand{\vec}[1]{\mathbf{#1}}
\newcommand{\uni}[1]{\,\text{#1}}
\newcolumntype{.}{D{.}{.}{-1}}

\url{} % www.pnas.org/cgi/doi/10.1073/pnas.0709640104}
\usepackage{url}
\begin{document}
\bibliographystyle{pnas}

\title{Energy-efficient neuromorphic classifiers}

\author{Daniel Mart\'{\i}\affil{1}{
  D\'{e}partement d'\'{E}tudes Cognitives, \'{E}cole Normale Sup\'{e}rieure - PSL Research University, Paris, France. Institut nationale de la sant\'{e} et de la recherche m\'{e}dicale, France.}\affil{3}{Center for Theoretical Neuroscience, Columbia University, New York, USA},
Mattia Rigotti\affil{2}{Physical Sciences Department, IBM T.\ J.\ Watson Research Center, Yorktown Heights, NY 10598}\affil{3},Mingoo Seok\affil{4}{Department of Electrical Engineering, Columbia University, New York, USA}
\and
Stefano Fusi\affil{3}{}}

% \contributor{Submitted to Proceedings of the National Academy of Sciences
% of the United States of America}

\maketitle

\begin{article}
\begin{abstract}
Neuromorphic engineering combines the architectural and computational principles of systems neuroscience with semiconductor electronics, with the aim of building efficient and compact devices that mimic the synaptic and neural machinery of the brain. Neuromorphic engineering promises extremely low energy consumptions, comparable to those of the nervous system. However, until now the neuromorphic approach has been restricted to relatively simple circuits and specialized functions, rendering elusive a direct comparison of their energy consumption to that used by conventional von Neumann digital machines solving real-world tasks. Here we show that a recent technology developed by IBM can be leveraged to realize neuromorphic circuits that operate as classifiers of complex real-world stimuli. These circuits emulate enough neurons to compete with state-of-the-art classifiers. We also show that the energy consumption of the IBM chip is typically 2 or more orders of magnitude lower than that of conventional digital machines when implementing classifiers with comparable performance. Moreover, the spike-based dynamics display a trade-off between integration time and accuracy, which naturally translates into algorithms that can be flexibly deployed for either fast and approximate classifications, or more accurate classifications at the mere expense of longer running times and higher energy costs. This work finally proves that the neuromorphic approach can be efficiently used in real-world applications and it has significant advantages over conventional digital devices when energy consumption is considered.	
\end{abstract}


\abbreviations{SVM: support vector machine |  SV: support vector | RCN: randomly connected neuron}

\subsection*{Introduction}

Recent developments in digital technology and machine learning are enabling computers to perform an increasing number of tasks that were once solely the domain of human expertise, such as recognizing a face in a picture or driving a car in city traffic. These are impressive achievements, but we should keep in mind that the human brain carries out tasks of such complexity using only a small fraction of the energy needed by conventional computers, the difference in energy consumption being often of several orders of magnitude. This suggests that one way to reduce energy consumption is to design machines whose architecture takes inspiration from the biological brain, an approach that was proposed by Carver Mead in the late 1980s~\cite{mead_1989} and that is now known as ``neuromorphic engineering''.  Mead's idea was to use very-large-scale integration (VLSI) technology to build electronic circuits that mimic the architecture of the nervous system. The first electronic devices inspired by this concept were analog circuits that exploited the subthreshold properties of transistors to emulate the biophysics of real neurons. Nowadays the term ``neuromorphic'' refers to any analog, digital, or hybrid VLSI system whose design principles are inspired by those of biological neural systems~\cite{indiveri_etal_fins2011}.

Neuromorphic hardware has convincingly demonstrated its potential for energy efficiency, as proven by devices that consume as little as a few picojoules per neural event (spike) \cite{livi_indiveri_ieecsymp2009,rangan_etal_ieecconf2010,chicca2014}. These devices contain however a relatively small number of elements (neurons and synapses) and they can typically perform only simple and specialized tasks, making it difficult to directly compare their energy consumption to that of conventional digital machines.

The situation has changed recently with the development by IBM of the TrueNorth processor, a neuromorphic device that implements enough artificial neurons to perform complex real-world tasks, like large-scale pattern classification~\cite{merolla_etal_science2014}. Here we show that a pattern classifier implemented on the IBM chip can achieve performances comparable to those of state-of-the-art conventional devices based on the von Neumann architecture. More importantly, our chip-implemented classifier uses 2 or more orders of magnitude less energy than current digital machines performing the same classification tasks. These results show for the first time the deployment of a neuromorphic device able to solve a complex task, while meeting the claims of energy efficiency contented by the neuromorphic engineering community for the last few decades.

\section*{Results}

\begin{figure*}[tbp]
  \centering
  \begin{minipage}[t]{17.7cm}
    \vspace{0pt}
    \includegraphics{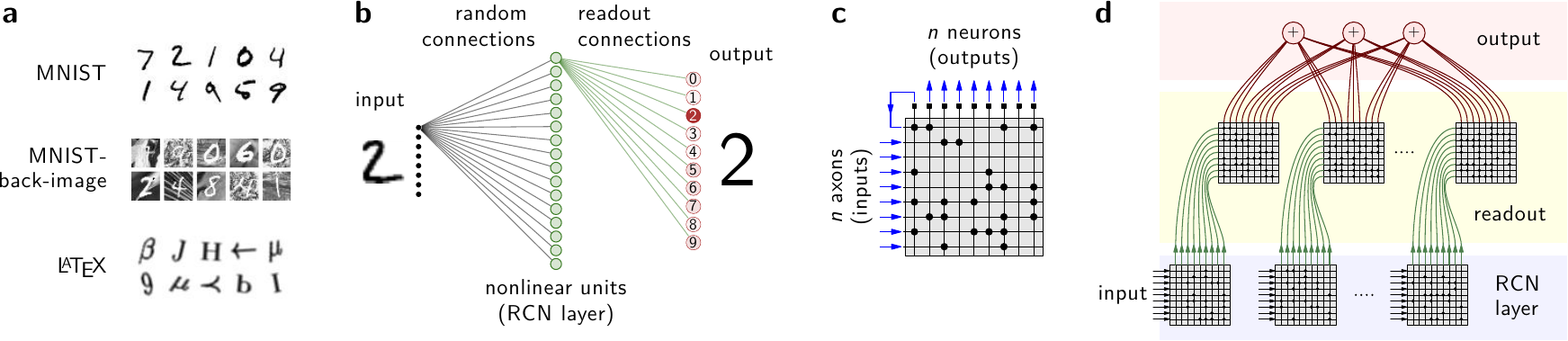}
  \end{minipage}
  \caption{\textbf{Datasets, architecture of the classifier, architecture of a single core, and chip
  implementation.} \textbf{a} Samples of the three datasets used to evaluate the performance of our classifier. MNIST contains handwritten digits (10 classes); MNIST-back-image contains the digits of the MNIST dataset on a background patch extracted randomly from a set of 20 images downloaded from the Internet; \LaTeX{} contains distorted versions of 293 characters used in the \LaTeX{} document preparation system. For more details about the datasets, see Methods. \textbf{b} Architecture of the neural network classifier. The images to classify are preprocessed (see Methods) and represented as patterns of activity of a population of $N_{\text{in}}$ input neurons (left, black dots). These input neurons send random projections to a second layer of $N$ \emph{Randomly Connected Neurons} (RCNs) (green circles), which transform nonlinearly their synaptic inputs into firing activity. The activity of the RCNs is then read out by a third layer of neurons, each of which is trained to respond to only one class (red circles).  \textbf{c} Architecture of a single core in the chip. Horizontal lines represent inputs, provided by the axons of neurons that project to the core. Vertical lines represent the dendrites of the neurons in the core (one dendrite per neuron). Active synapses are shown as dots in a particular axon-dendrite junction. The synaptic input collected by the dendrites is integrated and transduced into spike activity at the soma (filled squares on top). The spikes emitted by the neuron are sent via its axon to a particular input line, not necessarily on the same core. Blue lines represent the flow of input and output signals. The panel includes an example of internal connection: the upmost axon carries the output activity of the leftmost neuron in the core (other connections are left unspecified). \textbf{d} Implementation of the neural network classifier in a chip with connectivity constraints. The input is fed into all the cores in the RCN layer (shaded blue), whose neurons project to the input lines of readout cores (shaded yellow) in a one-to-one manner (green curves). The outputs of the readout units are combined together off-line to generate the response of the output neuron (shaded red). See the main text for the description of the different modules.
  \label{fig:basic}}
\end{figure*}

We chose pattern classification as an example of a complex task because of the availability of well-established benchmarks. A classifier takes an input, like the image of a handwritten character, and assigns it to one among a set of discrete classes, like the set of digits. To train and evaluate our classifiers we used three different datasets consisting of images of different complexity (see Fig.~\ref{fig:basic}a).

We start by describing the architecture of the classifier that we plan to implement on the neuromorphic chip. The classifier is a feed-forward neural network with three layers of neurons, and it can be simulated on a traditional digital computers. We will call this network the `neural classifier' to distinguish it from its final chip implementation, which requires adapting the architecture to the connectivity constraints imposed by the hardware. The neural classifier also differs from the final hardware implementation in that it employs neurons with a continuous activation function, whereas the IBM neuromorphic chip emulates spiking neurons. Despite the differences, the functionality of the neural classifier and its final chip implementation is approximately the same, as we show below. We list the procedure for adapting the architecture of the neural classifier into its chip implementation as a contribution in its own right, since it can be directly extended for the implementation of generic neural systems on other hardware substrates.

\subsection*{Architecture of the neural classifier}

Figure~\ref{fig:basic}b illustrates the three-layer neural classifier. The first layer encodes the preprocessed input and projects to the neurons in the intermediate layer through connections with random weights. Each of these \emph{Randomly Connected Neurons} (RCNs) receives therefore a synaptic current given by a randomly weighted sum of the inputs, which the RCNs transform into activation levels in a non-linear way---in our case, through a linear rectification function: $f(x) = x$ if $x>0$, and $0$ otherwise. The combination of a random mixing of the inputs together with a non-linear input-output transformation efficiently expands the dimensionality of the resulting signal (see e.g.\ \cite{Jaeger2004,Buonomano2009,barak_etal_jns2013}), thereby increasing the chances that downstream neurons can discriminate signals belonging to distinct classes. This discrimination is carried out by a set of \emph{output units} in the last layer, which compute a weighted sum of the RCNs activity. The weights are trained so that each output unit responds to one separate class (\emph{one-vs-all} code). Details are given in the Methods. Once the network is trained, a class is assigned to each input patterns according to which output unit exhibits the highest activation.

\subsection*{Chip implementation of the neural classifier}

We implemented the neural classifier on the IBM neuromorphic chip described in \cite{arthur_etal_ijcnn2012,merolla_etal_science2014}. The first step of the conversion of the abstract neural classifier to an explicit chip implementation is the transformation of the input patterns into a format that is compatible with the spike-based coding of the TrueNorth system. For this we simply employ a firing rate coding and convert the integer value of every input component to a spike train with a proportional number of spikes, a prescription that is commonly used in neurocomputational models such as the Neural Engineering framework \cite{Eliasmith2012}.
Specifically, input patterns are preprocessed and formatted into 256-dimensional vectors representing the firing activity of the input layer (the same preprocessing step was applied in the neural classifier, see Methods). This vector of activities is then used to generate 256 regular-firing spike trains that are fed into a set of cores with random and sparse connectivity. This set of cores constitutes the RCN layer. Like in the neural classifier, the neurons in the RCN layer receive synaptic inputs that consist of randomly weighted combinations of the input, and transform their synaptic inputs into firing activity according to a nonlinear function. On the chip this function is given by the neuronal current-to-rate transduction, which approximates a linear-rectification function~\cite{fusi_mattia_neco1999}.

Discriminating the inputs coming from the RCN layer requires each output unit to read from the whole layer of RCNs, which in our implementation contains a number of neurons $N$ that can be as large as $2^{14}$. Moreover, all the readout connectitions have to be set at the weights computed by the training procedure. These requirements exceed the constraints set by the chip design, in terms of the maximal number of both incoming and outgoing connections per neuron, as well as the resolution and the freedom with which synaptic weights can be set. In this paragraph we will present a set of prescriptions that will allow us to circumvent these limitations, and successfully instantiate our neural classifier on the IBM system. The prescriptions we are presenting are specific to the TrueNorth architecture, but the types of constraints that they solve are shared by any physical implementation of neural systems, whether it is biological or electronic. It is therefore instructive to discuss the constraints and the prescriptions to obviate them in detail, as they can be easily extended to other more generic settings.

\vspace{2mm}

\noindent {\sl 1. Constraints on connectivity.}
The IBM chip is organized in cores, each of which contains 256 integrate-and-fire neurons and 256 input lines that intersect with one another forming a crossbar matrix of programmable synapses (Fig.~\ref{fig:basic}c). Each neuron can connect to other neurons by projecting its axon (output) to a single input line, either on the same core or on a different core. With this hardware design the maximum number of incoming connections per neuron, or \emph{fan in}, is 256. Likewise, the maximal number of outgoing connections per neuron, or \emph{fan out}, is 256, each of which are restricted to target neurons within a single core.

\vspace{2mm}

\noindent {\sl 2. Constraints on synaptic weight precision.}
Synapses can be either inactive or active. The weight of an active synapse can be selected from a set of four values given by signed integers of 9-bit precision. These values can differ from neuron to neuron. Which of the four values is assigned to an active synapse depends on the input line: all synapses on the same input line are assigned an index that determines which of the four values is taken by each synapse (e.g.\ if the index assigned to the input line is 2, all synapses on the input line take the second value of the set of four available synaptic weights, which may differ from neuron to neuron).

The design constraints that we just described can be overcome with the following set of architectural prescriptions.

\vspace{2mm}

\noindent {\sl P1. Overcoming the constraints on connectivity.} We introduced an intermediate layer of neurons, each of which integrates the inputs from 256 out of the total $N$ RCNs. Accordingly, the firing rates of these intermediate neurons represent a $256/N$ portion of the total input to an output unit. These partial inputs can then be combined by a downstream neuron, which will have the same activity as the original output unit. If the total number of the partial inputs is larger than the total number of incoming connections of the neurons that represent the output units (in our case 256), the procedure can be iterated by introducing additional intermediate layers. The final tree will contain a number of layers that scales only logarithmically with the total number of RCNs. For simplicity we did not implement this tree on chip and we summed off-chip the partial inputs represented by the firing activity of the readout neurons.
Notice also that this configuration requires readout neurons to respond approximately linearly to their inputs, which can be easily achieved by tuning readout neurons to operate in the linear regime of their current-to-rate transduction function (i.e., the regime in which their average input current is positive). This procedure strongly relies on the assumption that information is encoded in the firing rates of neurons; if the spiking inputs happen to be highly synchronized and synchronization encodes important information, this approach would not work.

\vspace{2mm}

\noindent {\sl P2. Overcoming the constraints on synaptic weight precision.} Reducing the weight precision after learning usually only causes moderate drops in classification performance. For example, in the case of random uncorrelated inputs, the scaling properties of the capacity of the classifier (i.e., number of classes that can be correctly classified) remain unchanged, even when the number of states of the synaptic weights is reduced to two \cite{sompolinsky_physA1986}. Instead, the performance drop is catastrophically larger when the weight precision is limited also during learning \cite{amitfusi94,fusi2002} and in some situations the learning problem becomes NP-complete \cite{garey_johnson1979}
In our case the readout weights are determined off-chip, using digital conventional computers that operate on 64 bit numbers, and then quantized in the chip implementation. The performance drop is almost negligible for a sufficient number of synaptic levels. In our case we quantized the readout weights of the original classifier on an integer scale between $-28$ and $28$. Each quantized weight was then implemented as the sum of four groups of 6 synaptic contacts, where each contact in the group can either be inactive (value 0) or activated at one of the 6 values: $\pm 1,\pm 2,\pm 4$. The multiplicity of this decomposition ($19$ can be for instance decomposed as $(4)+(1+4)+(1+4)+(1+4)$ or $(2)+(2+4)+(2+4)+(1+4)$) is resolved by choosing the decomposition that is closest to a balanced assignment of the weights across the 4 groups (e.g.\ $19=(4)+(1+4)+(1+4)+(1+4)$). This strategy requires that each original synapse be represented by 24 synapses. We implemented this strategy by replicating each readout neuron 24 times and by distributing each original weight across 24 different dendritic trees. These synaptic inputs are then summed together by the off-line summation of all readout neuron activities that correspond to the partial inputs to a specific output unit (see Methods for details). A similar strategy can be used to implement networks with synaptic weights that have even a larger number of levels and the number of additional synapses would scale only logarithmically with the total number of synaptic levels that is required. However, it is crucial to limit individual synapses to low values, in order to avoid synchronization between neurons. This is why we limited to 4 the maximum synaptic value of individual synapses of the chip.

  % \begin{figure*}[tbp]
    % \centering
    % \includegraphics{figures/accumulated_spikes_0002_detrended}\qquad
    % \includegraphics{figures/accumulated_spikes_0010_detrended}
    % % \includegraphics{figures/rastergram_0002}
    % \caption{\textbf{Example of correct classifications computed by the chip.}\: \textbf{a}, \textbf{b}, Number of emitted readout spikes during an easy (a) and a difficult (b) classification, after removing the trend caused by the intrinsic constant currents. The chip uses $N=8192$ \RCN{s}. Each curve corresponds to the readout output associated with the digit indicated by color code.}
    % \label{fig:patrec_ex}
  % \end{figure*}

\subsection*{Classification performance and speed-accuracy trade-off}

\begin{figure}[tbp]
   \centering
   \begin{minipage}[t]{8.5cm}
     \includegraphics{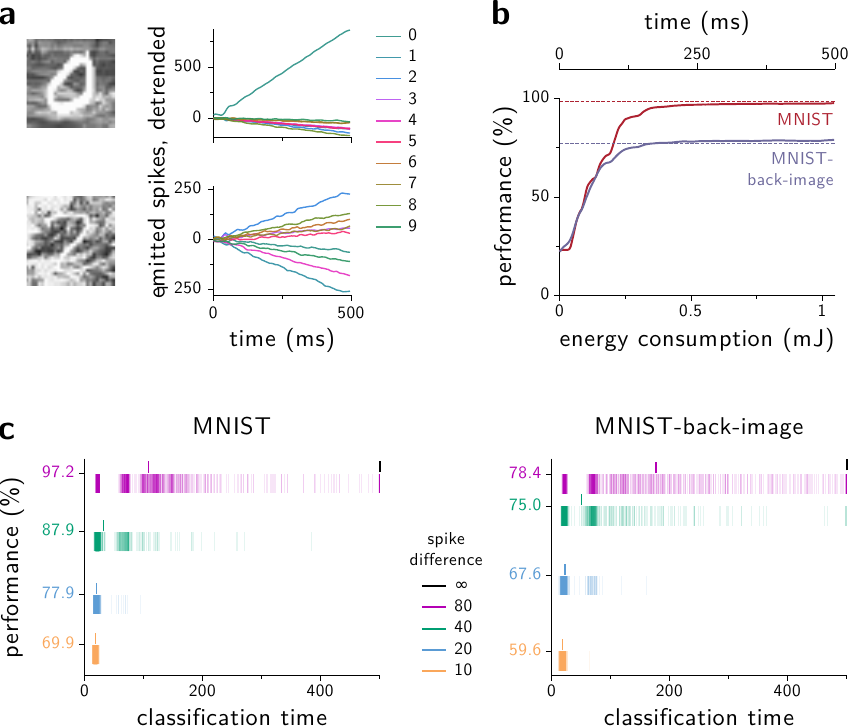}
   \end{minipage}
   \caption{\textbf{The neuromorphic classifier in action.} \textbf{a} Spikes emitted by readout neurons during an easy (top) and a difficult (bottom) classification, after removing the trend caused by the intrinsic constant currents. Each curve corresponds to the readout output associated with the digit indicated by the color code. Samples are drawn from the MNIST-back-image dataset. \textbf{b} Test error as a function of classification time (i.e., the time over which spikes are integrated) and energy. The error is averaged over the first 1000 test samples of the MNIST (red) and MNIST-back-image (blue) datasets. Each dashed horizontal line indicates the best test error achieved with support vector classifiers for a given dataset, based on the evaluation of the whole test set. \textbf{c} Classification times for different thresholds in spike difference (as indicated in the legend), for the MNIST and MNIST-back-image datasets. For each threshold we plot all classification times (thin lines) as well as the sample mean (shorter ticks on top). The performances associated with each threshold are indicated in the $y$-axis. When the threshold in spike difference is infinite (black), the classification is assessed at $t=500\uni{ms}$ (i.e., there is no stopping criterion). In all panels the chip uses $N=16384$ RCNs. \label{fig:chip_dyn}}
  \end{figure}

Our \emph{neuromorphic classifier} implemented on the TrueNorth chip was emulated on a simulator developed by IBM. As the TrueNorth chip is entirely digital, the simulator reproduces exactly the behavior of the chip \cite{arthur_etal_ijcnn2012}. In Fig.~\ref{fig:chip_dyn}a we show the dynamics of two typical runs of the simulator classifying images from the MNIST-back-image dataset. Upon image presentation, the RCNs in the intermediate layer start integrating the input signal (not shown) and, a few tens of milliseconds later, they start emitting spikes, which are passed  to the readout neurons. The figure shows the total number of spikes emitted by the readout neurons since input activation, after subtracting the overall activity trend caused by baseline activity.

For simple classifications, in which the input is easily recognizable, the readout neuron associated with the correct class is activated in less than $100\,\text{ms}$ (Fig.~\ref{fig:chip_dyn}a, top). More difficult cases require the integration of spikes over longer time intervals, as the average synaptic inputs to different readout neurons can be very similar (Fig.~\ref{fig:chip_dyn}a, bottom). This suggests that the performance of the classifier, as measured by the classification error rate on the test set, should improve with longer integration intervals. This trade-off between speed and performance is illustrated in Fig.~\ref{fig:chip_dyn}b, which shows the classification performance versus elapsed time for the MNIST and MNIST-back-image datasets. The performance increases monotonically with time until it saturates in about half a second, with a highest performance of $97.27\%$ for MNIST ($98.2\%$ with 10-fold bagging), and $77.30\%$ for MNIST-back-image. These performances are not too far from the best classification results achieved so far: $99.06\%$ for MNIST (using maxout networks on the permutation invariant version of the MNIST dataset, which does not exploit any prior knowledge about the two-dimensional structure of the patterns~\cite{goodfellow2013}) and $77.39\%$ for MNIST-back-image (with support vector classifiers~\cite{cho2010}, although methods combining deep nets, feature learning, and feature selection can achieve performances as high as $87.75\%$~\cite{sohn_etal_icml2013}).

\subsection*{Energy-speed-accuracy trade-off}
As just discussed, accuracy has a cost in term of energy because longer integration times entail more emitted spikes per classification and a larger baseline energy costs, which in our case is the dominant contribution to the total energy consumption. We estimated the energy consumption as described in section \ref{sec:ibmchipenergy} and we found that the energy per classification never exceeds $1\uni{mJ}$ for our network configuration. With the energy needed to keep lit a $100\uni{W}$ light bulb for a second, one could perform $10^5$ classifications, which is equivalent to around one classification per second uninterruptedly for almost one day. Notice that this estimate is based on a classification that lasts $0.5\uni{s}$ and, therefore, does not take into account the fact that most patterns are correctly classified in a significantly shorter time (see Fig.~\ref{fig:chip_dyn}a, top). If the integration and emission of spikes is stopped as soon as one of the output units is significantly more active than the others, then the average energy consumption can be strongly reduced. The criterion we used to decide when to stop the integration of spikes (and thus the classification) was based on the spikes emitted by the readout units. Specifically, we monitored the cumulative activity of each output unit by counting all the spikes emitted by the corresponding readout neurons. We stopped the classification when the accumulated activity of the leading unit exceeded that of the second unit by some threshold. The decision was the class associated with the leading output unit.

In Fig.~\ref{fig:chip_dyn}c we show the performances and the corresponding classification times for several thresholds. Low thresholds allow for faster yet less accurate classifications. In both the MNIST and MNIST-back-image datasets, the patterns that require long classifications times are rare. While the performance barely changes for large enough thresholds, the average classification time can be substantially reduced by lowering the threshold. For example, for the MNIST dataset the classification time drops by a factor of 5 (from $500\uni{ms}$ to $100\uni{ms}$) and, accordingly, so does the energy consumption (from $1\uni{mJ}$ to $0.2\uni{mJ}$). Faster classifications are also possible by increasing either the average firing rate or the total number of RCNs, both of which entail an increase in energy consumption, which might be partially or entirely compensated by the decrease in the classification time. These expedients will speed up the integration of spike-counts and, as a result, the output class will be determined faster.

In all cases both the energy cost and the classification performance increase with the total number of emitted spikes or, equivalently, with integration time, if the average firing rate is fixed. This is a simple form of a more general energy-speed-accuracy trade-off, a phenomenon that has been described in several biological information-processing systems (e.g.\ \cite{lan2012}), and that can confer great functional flexibility to our classifier. One advantage of basing the computation on a temporal accumulation of spikes is that the classifier can be interrupted at any time at the cost of reduced performance, but without compromising its function. This is in stark contrast to some conventional clock-based centralized architectures whose mode of computation crucially relies on the completion of entire monolithic sets of instructions. We can then envisage utilization scenarios where a spiking-based chip implementation of our classifier is required to flexibly switch between precise long-latency classifications (like, e.g., those involving the identification of targets of interest) and rapid responses of limited accuracy (like the quick avoidance of imminent danger).

%In the remaining of the article we will be working with fixed classification times ($500\,\text{ms}$), because for such duration the chip already compares favorably to other devices when energy consumption is considered.

Notice that both the simulated and implemented networks, although entirely feed-forward, exhibit complex dynamics leading to classification times that depend on the difficulty associated with the input. This is because neurons are spiking and the final decision requires some sort of accumulation of evidence. When a stimulus is ambiguous, the units representing the different decisions receive similar inputs and the competition becomes harder and longer. This type of behavior is also observed in human brains \cite{tang_etal_neuron2014}.

We will now focus on the comparison of energy consumption and performance between the neuromorphic classifier and more conventional digital machines.

\subsection*{Energy consumption and performance: comparison with conventional digital machines}
We compared both the classification performance and the energy consumption of our neuromorphic classifier to those obtained with conventional digital machines implementing Support Vector Machines (SVMs). SVMs offer a reasonable comparison because they are among the most successful and widespread techniques for solving machine-learning problems involving classification and regression \cite{boser1992,cortes_vapnik_ml1995,vapnik1997}, and because they can be efficiently implemented on digital machines.

To better understand how the energy consumption scales with the complexity of the classification problem, it is useful to summarize how SVMs work. After training, SVMs classify an input pattern according to its similarity to a set of templates, called the support vectors, which are determined by the learning algorithm to define the boundaries between classes. The similarity is expressed in terms of the scalar product between the input vectors and the support vectors. As argued above, we can improve classification performance by embedding the input vectors in a higher-dimensional space before classifying them. In this case SVMs evaluate similarities by computing classical scalar products in the higher-dimensional space. One of the appealing properties of SVMs is that there is no need to compute explicitly the transformation of inputs into high-dimensional representations. Indeed, one can skip this step and compute directly the scalar product between the transformed vectors and templates, provided that one knows how the distances are distorted by the transformation. This is known as the ``kernel trick'' because the similarities in a high-dimensional space can be computed and optimized over with a kernel function applied to the inputs. Interestingly, the kernel associated with the transformation induced by the RCNs of our neural classifier can be computed explicitly in the limit of a large number of RCNs \cite{cho2010}. This is also the kernel that we used to compare the performance of SVMs against that of our neural classifier.

Unfortunately, classifying a test input by computing its similarity to all support vectors becomes unwieldy and computationally inefficient for large datasets, as the number of support vectors typically scales linearly with the size of the training set in many estimation problems \cite{steinwart2008}. This means that the number of operations to perform, and hence the energy consumption per classification, also scales with the size of the training set. This makes SVMs and kernel methods computationally and energetically expensive in many large-scale tasks. In contrast, our neural network algorithm evaluates a test sample by means of the transformation carried out by the RCNs. If the RCN layer comprises $N$ neurons and the input dimension is $N_{\text{in}}$, evaluating the output of a test sample requires $O(N_{\text{in}} \cdot N)$ synaptic events. Thus for large sample sizes, evaluating a test sample in the network requires far fewer operations than when using the ``kernel trick'', because the number is  effectively independent of the size of training set (cfr.~\cite{rahimi2008,le2013}). Systems such as ours may therefore display considerable energy advantages over SVMs when datasets are large.

\begin{figure*}[tbp]
   \centering
   \begin{minipage}[t]{14cm}
     \includegraphics{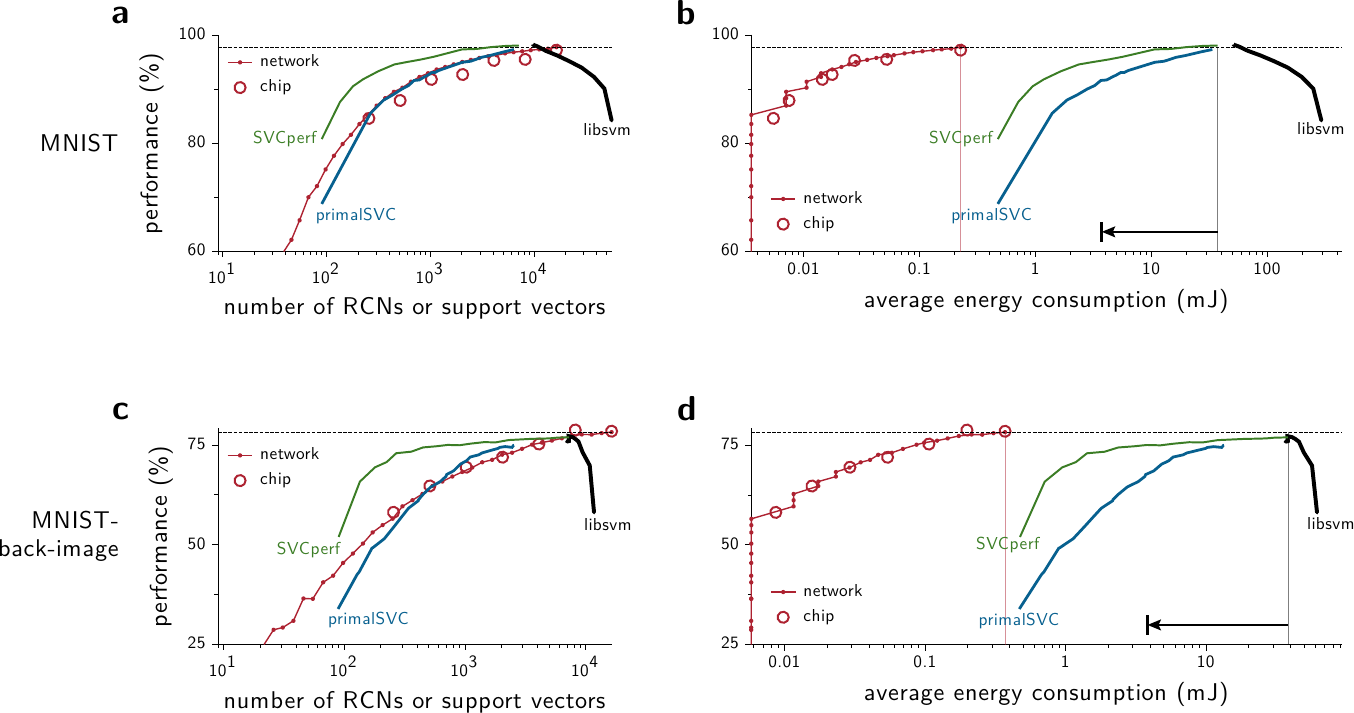}
   \end{minipage}
   \caption{\textbf{Energy-accuracy trade-off.} \textbf{a,c} Dependence of the classification accuracy on the number of Randomly Connected Neurons (RCNs) in the neural classifier and on the number of support vectors (SVs) in the SVC. Panel \textbf{a} shows this dependence for the MNIST dataset, and panel \textbf{b} for the MNIST-back-image. As the number of RCNs increases, the classifier becomes more accurate at the cost of higher energy consumptions (\textbf{b,d}). The energy consumption is based on the average time it takes to the neural classifier to perform the classification (see Fig.~\ref{fig:chip_dyn}c). We also show the performance achieved by three different implementations of support vector classifiers (legend code: SVC, \texttt{libsvm}; rSVC, reduced primal; SVC; SVCperf, cutting plane subspace pursuit). The algorithms rSVC, SVCperf minimize the number of support vectors (SVs) with respect to the optimal value and reduce, therefore, the energy consumption levels at test time. The number of SVs used by the standard algorithm (libsvm), on the other hand, can go beyond the optimal value by reducing sufficiently the soft-margin parameter and pushing the classifier to overfit the data. In all cases, the energy consumption increases linearly with the number of SVs, as the number of operations per classification at test time scales linearly with the number of SVs. The vertical thin lines indicate the abscissa at best performance for the IBM chip (red) and SVM implementations (black). For reference we indicate the best performance achieved by the chip with a horizontal dashed line. The horizontal arrow indicates the reduction in energy consumption that would be attained if the efficiency of digital machines reached the theoretical lower bound estimated by~\cite{hasler_marr_fins2013}. The relation between number of  SVs and energy consumption was determined by simulating the i7 Intel chip running a program that implements an SVM at test time. \textbf{b} Same as \textbf{a}, but on the MNIST-back-image dataset. In both cases our neuromorphic classifier exhibits an energy cost per classification that is orders of magnitude smaller.
   \label{fig:resources}}
\end{figure*}

\begin{figure*}[tbp]
   \centering
   \begin{minipage}[t]{14.7cm}
     \includegraphics{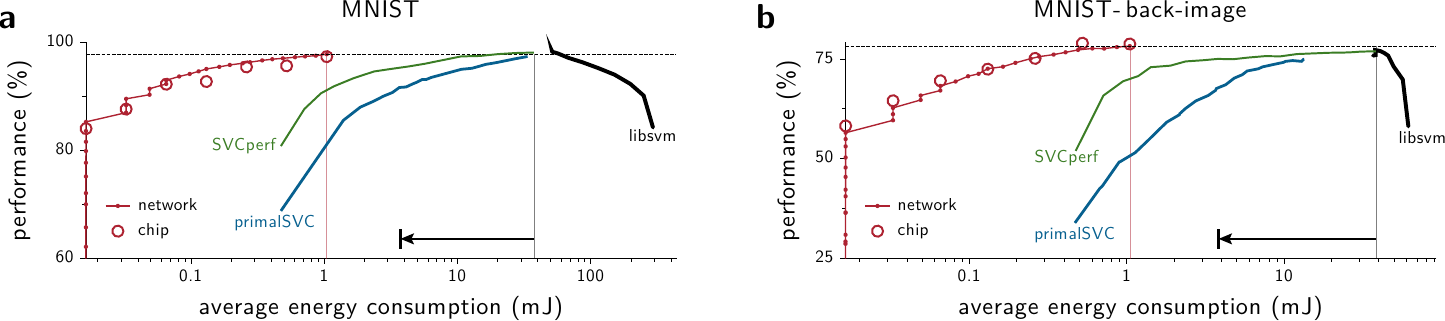}
   \end{minipage}
   \caption{\textbf{Performance versus energy consumption at a fixed classification time.} Panels are like in Figs.~\ref{fig:resources}b,d, classification time is now fixed at $500\uni{ms}$, rather than determined by a stopping criterion.
   \label{fig:resources_fixedT}}
\end{figure*}

In Fig.~\ref{fig:resources} we compare the energy consumption and performance of the neuromorphic classifier to those of an SVM implemented on a conventional digital machine. More specifically, we estimated the energy expenditure of a digital SVM using a simulator of the Intel i7 processor, which was the machine with the best energy performance among those that we simulated (see Methods section~\ref{sec:digital_machines} and Discussion). The energy cost per support vector per pattern was estimated to be around $5.2\,\mu\text{J}$,  a quantity that is not far above what is considered as a lower bound on energy consumption for digital machines~\cite{hasler_marr_fins2013}. For both the neuromorphic classifier and the digital SVM we progressively increased the performance of the classifiers by increasing the number of RCNs (in the case of the neuromorphic classifier), and by varying the number of support vectors (in the case of the SVM), see Figures \ref{fig:consumption}a,c. For the SVM we tried three different algorithms to minimize the number of support vectors and hence the energy consumption (for more details, see caption of Fig.~\ref{fig:resources} and Methods). For the IBM chip we estimated the energy consumption both in the case in which we stopped the classifications with the criterion described in the previous section and in the case in which the classification time was fixed at $500\uni{ms}$ (see Fig.~\ref{fig:resources_fixedT} in Suppl.\ Info.). In both cases the energy consumption is significantly lower for the neuromorphic classifier, being in the former case approximately 2 orders of magnitude smaller for both the MNIST and the MNIST-back-image datasets, while still achieving comparable maximal performances (Fig.~\ref{fig:consumption}b,d).

\subsection*{Scalability}
The MNIST dataset only has 10 output classes. We wondered whether the advantage of the neuromorphic classifier in terms of energy consumption is preserved when the number of classes increases and the classification task becomes more complex. To study how the energy consumption scales with the number of classes we used the \LaTeX{} dataset, which contains 293 classes of distorted characters. We progressively increased the number of classes to be learned and classified and we studied the performance and the energy consumption of both the digital implementation of the SVM and the neuromorphic classifier. Specifically, given a number of classes that was varied between 2 and 293, we selected a random subset of all the available classes, and we trained both the SVM and the neural classifier on the same subset. The results are averaged over 10 repetitions, each one with a different sample of output classes.

To make a meaningful comparison between the the energy consumed by a SVM and the neuromorphic classifier, we equalized all the classification accuracies, as follows. For each classification problem we varied the margin penalty parameter of the standard SVC using grid search and picked the best performance achieved. We then varied the relevant parameters of the other two classifiers so that their classification accuracy matched or exceeded the accuracy of the standard SVC. Specifically, we progressively increased the number of basis functions (in the primalSVC method) and the number of RCNs (in the neural classifier) until both reached the target performance. For each classification problem we averaged over 10 realizations of the random projections of the neural classifier.

\begin{figure*}[tbp]
  \centering
  \includegraphics{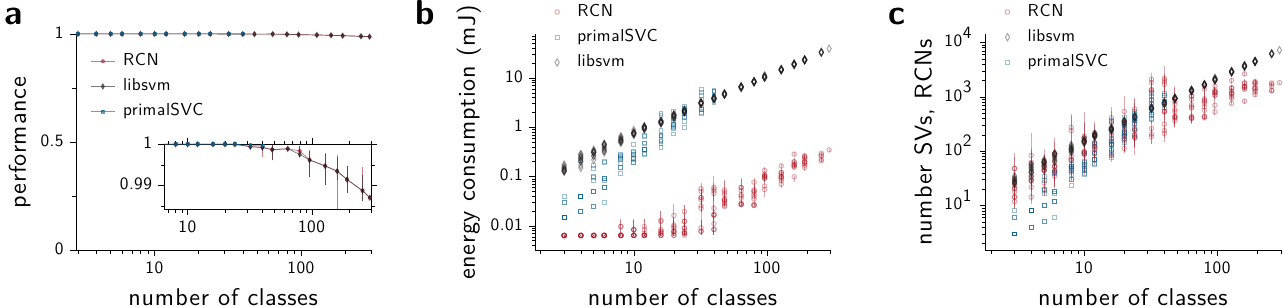}
  \caption{\textbf{Dependence of the energy consumption on the number of classes}\: \textbf{a} Classification accuracy for the neural classifier and for two SVM algorithms, as a function of the number $C$ of classes for the \LaTeX{} dataset. The parameters of the different classifiers are tuned to have approximately the same classification accuracy. \textbf{b} Energy consumption as a function of the number of classes, for the \LaTeX{} dataset. Given a number $C$ of classes, every point in the plot is obtained by training a given classifier on $C$ randomly sampled classes among the 293 available ones. This procedure is repeated 10 times for every value of $C$ and every type of classifier. Each datapoint associated with the neural classifier ('RCN') was in turn estimated from a sample of 10 realizations of the random connections (squares indicate sample means, errorbars indicate the 0.1 and 0.9 fractile of the sample). \textbf{c} As in \textbf{b}, but number of support vectors and RCNs as a function of the number of classes. \label{fig:consumption}}
\end{figure*}

The results are summarized in Fig.~\ref{fig:consumption}.
%For both the SVM implemented on a von-Neumann digital machine and the neuromorphic classifier, the energy grows approximately linearly with the number of classes, if the number of classes is large enough. 
The energy consumption is about two orders of magnitude larger for the SVM throughout the entire range of variation of the number of classes that we considered, although for a very small (2--3) number of classes the advantage of the neuromorphic classifier strongly reduces, most likely because the algorithms to minimize the number of SVs work best when the number of classes is low.  This plot indicates that the energy advantage of the neuromorphic classifier over SVMs implemented on conventional digital machines is maintained also for more complex tasks involving a larger number of classes.

It is interesting to discuss the expected scaling for growing number of classes. Consider the case of generic $C$ classes multi-class problems solved through reduction with multiple combined binary SVMs. In a one-vs-all reduction scheme, each binary classifier is trained to respond to exactly one of the $C$ classes, and hence $C$ SVMs are required. For each SVM, one needs to compute the scalar products between the test sample to be classified and the $N_{\text{SV}}$ support vectors. Each scalar product requires $N_{\text{in}}$ multiplications and sums.
In the favorable case in which all binary classifiers happen to share the same support vectors, the scalar products can be computed only once and would require $N_{\text{in}} \cdot N_{\text{SV}}$ operations. These $N_{\text{SV}}$ scalar products then need to be multiplied by the corresponding coefficients, which are different for the different SVMs. This requires additional $C N_{\text{SV}}$ operations. If $N_{\text{SV}}$ scales linearly with $C$, as in the cases we analyzed, then the total energy $E$ will scale as
\begin{equation*}
  E \sim N_{\text{in}} C + C^2.
\end{equation*}
When $C$ is small compared to $N_{\text{in}}$, the first term dominates, and the expected scaling is linear. However, for $C > N_{\text{in}}$ the scaling is expected to be at least quadratic. It can grow more rapidly if the support vectors are different for different classifiers.

Interestingly the expected scaling for the neural network classifiers that we considered is the same. The energy consumption mostly depends on the number of needed cores. This number will be proportional to the number of RCNs, $N$, multiplied by the number of classes. Indeed, each core can receive up to 256 inputs, so the total number of needed cores will be proportional to $\lceil N/256 \rceil$, with $\lceil \cdot \rceil$ denoting the ceiling function. Moreover, the number of readout units, which are the output lines of these cores, will be proportional to the number of classes. Hence the $NC$ dependence. In the cases we analyzed $N$ depends linearly on the number of classes, and hence the energy depends quadratically on $C$, as in the case of the SVMs when $C$ is large enough. Notice that the there is a second term which also scales quadratically with $C$ that contributes to the energy. The second term comes from the necessity of replicating the RCNs $C$ times, due to the limited fan out of the RCNs. Again, under the assumption of $N\sim C$, also this term will scale quadratically with $C$.

Given that the scaling with the number of classes is basically the same for the neuromorhic classifier as for the SVMs, it is not unreasonable to hypothesize that the energy consumption advantage of the neuromorphic implementation would be preserved also for a much larger number of classes.

% \begin{figure}[H]
  % \centering
  % \includegraphics{figures/nsvs_vs_nclasses_optimal_means}
  % \caption{\textbf{Dependence of performance on the number of classes}\: This is good but it can be combined together with figure 5. For symmetry I would also plot the number  of RCNs vs number of classes. May be everything can be plotted in the same graph (I guess the relation between energy consumption and RCNs or SVs is linear).
  % Number of support vectors as a function of the number of classes, for the \LaTeX{} dataset. For a given number $C$ of classes, we chose randomly 10 of the ${293 \choose C}$ possible class combinations, and for each such combination we trained a support vector classifier. Each data point represents the sample mean of the 10 realizations; error bars represent the associated sample standard deviation.}
  % \label{fig:svc}
% \end{figure}

%% \begin{table*}
%%  \small
%%    \centering
%%    \begin{tabular}{lcc}
%%      \toprule
%%      Method & Test error \\
%%      \midrule
%%      Pseudoinverse with 8192 \RCN{s} & 23.734\% (50000 samples)\\
%%      After quantizing, with $n_{\text{w}}=64$ and $J_{\text{max}}=64$ & 23.732\% (50000 samples)\\
%%                                            \hfill $J_{\text{max}}=32$ & 23.944\% (50000 samples)\\
%%      Chip & 22.25\% (2000 samples)\\
%%      \bottomrule
%%    \end{tabular}
%%    \caption{Results}
%%    \label{tab:performances}
%%  \end{table*}
%%

\section{Discussion}

Our results indicate that neuromorphic devices are mature enough to achieve performances on a real-world machine-learning task that are comparable to those of state-of-the-art conventional devices with von Neumann architecture, all just by using a tiny fraction of their energy. Our conclusions are based on a few significant tests, based on a comparison limited to our neuromorphic classifier and a few digital implementations of SVMs. This clearly restricts the generality of our results and does not preclude situations in which the advantage of the neuromorphic approach might be less prominent. In any case, the merit of our study is to offer a solid comparison with implementations on current conventional digital platforms that are energy-efficient themselves. In particular, the algorithm we used on conventional digital machines involves only multiplications between matrices and vectors, the efficiency of which has been dramatically increased in the last decades thanks to optimized parallelization. Furthermore, not only we tried to match the classification performance of the competitors, but we also considered two additional SVM algorithms that minimize the number of support vectors, and hence the final number of operations. Other choices for SVM algorithms would certainly lead to different estimates for energy consumption, but it is rather unlikely that they would change across 2 orders of magnitude. It is possible that full custom unconventional digital machines based, e.g.,\ on field programmable gate arrays (FPGAs) would be more energy-efficient, but it is hard to imagine that they would break the predicted energy wall discussed in \cite{hasler_marr_fins2013}. If this assumption is right, neuromorphic hardware would always be more efficient when performing the type of tasks that we considered. Moreover, analog neuromorphic VLSI or unreliable digital technologies might allow for a further reduction of energy consumption, probably by another order of magnitude \cite{chicca2014,arthur2011,han_orshanksky_ieee2013}. The current energy consumption levels achieved by analog systems are very close to those of biological brains in terms of energy per spike, although many of these systems are relatively small and it is unclear whether they can ever be extended to brain-scale architectures.

Other custom chips that can solve real-world tasks have been designed. An example is the FPGA chip NeuFlow, designed to implement convolutional networks for visual recognition. The chip is digital and uses as little as $4.9\times 10^{11}\,\text{operations} / \text{W}$ or, equivalently, $2\,\text{pJ} / \text{operation}$.

It is also interesting to discuss the performance of other conventional digital processors in the benchmarks we examined. Let us consider for example the implementations of SVMs classifying the MNIST digits with about $10^4$ support vectors, which is roughly the number of vectors we need to achieve the best classification accuracy. As we have shown, the Intel i7 takes about $10\uni{ms}$ to perform a classification, at an approximate cost of $50 \uni{mJ}$. The IBM chip, in contrast, required $1 \uni{mJ}$ for the longest classification times ($500\uni{ms}$), and $0.2 \uni{mJ}$ for the average classification time ($100\uni{ms}$). We also quantified the energy cost of the ARMv7, which is a more energy-efficient yet slower microprocessor often used in mobile technologies. Its energy consumption per classification was substantially higher, around $700 \uni{mJ}$. The main reason for this high consumption is that it takes more than $0.6$ seconds to perform a single classification. And the baseline consumption, which increases linearly with the classification time, is a large portion of the total energy needed for a classification. Finally, we considered the recent Xeon Phi, which has a massively parallel architecture and is employed in high performance computing applications. As we do not have a simulator for the Phi, we could only indirectly estimate a lower bound for the energy consumption (see Methods for more details). According to our estimate, a single classification requires only $0.2\,\mu\text{s}$ and uses about $16\uni{mJ}$, which would be significantly lower than the energy cost of the i7 and very close to the estimated lower limit of energy consumption~\cite{hasler_marr_fins2013}, but still larger than the consumption of the IBM chip. Notice however that both the classification time and the energy consumption of the Xeon Phi processor are very likely to be grossly underestimated, as they are simply derived from the peak performance of $100\uni{Tflop/s}$. The estimates for the i7 and the ARMv7 are significantly more reliable, because we derived them by simulating the processors.

To summarize, our results compellingly suggest that the neuromorphic approach is finally competitive in terms of energy consumption in useful real-world machine learning tasks and constitutes a promising direction for future scalable technologies. The recent success of deep networks for large-scale machine learning \cite{krizhevsky2012,deng2013} makes neuromorphic approaches particularly relevant and valuable. This will be certainly true for neuromorphic systems with synaptic plasticity, which will enable these devices to learn autonomously from experience. Learning is now available only in small neuromorphic systems \cite{mitra2009, giulioni2011, arthur2006}, but hopefully new VLSI technologies will allow us to implement it also in large-scale neural systems.

\begin{acknowledgments}
This work was supported by DARPA SyNAPSE, Gatsby Charitable Foundation, Swartz Foundation and Kavli Foundation. We are grateful to the IBM team led by Dr.~D.~Modha for their assistance with the IBM chip simulator. In particular we thank John Arthur and Paul Merolla for their help with the estimate of the chip energy consumption. DM acknowledges the support from the FP7 Marie Curie Actions of the European Commission and the ANR-10-LABX-0087 IEC and ANR-10-IDEX-0001-02 PSL grants.
\end{acknowledgments}

\section*{Materials and Methods}

\subsection*{Images sets for classification benchmarks}
We used three datasets in our study: MNIST, MNIST-back-image, and \LaTeX{}. The MNIST dataset consists of images of handrwitten digits (10 classes)~\cite{LeCun1998}. The MNIST-back-image dataset contains the same digits of MNIST, but in this case the background of each pattern is a random patch extracted from a set of 20 black and white images downloaded from the Internet~\cite{larochelle_etal_procicml2007}. Patches with low pixel variance (i.e.\ containing little texture) are discarded. The \LaTeX{} dataset consists of distorted versions of 293 characters used in the \LaTeX{} document preparation system~\cite{amit_geman_neco1997,amit_2d_2002}. All datasets consist of $l\times l$ pixel gray-scale images, and each of such pixel images is associated with one out of $C$ possible classes. The size of the pixel images, the number of classes, and the sizes of the training and test sets depend on the data set (see table below).
  \begin{table}[tbp]
    \centering
    \small
    \begin{tabular}{ccccc}
      \toprule
      \small\small Dataset & \parbox{0.9cm}{\centering \small image\\ size} & \parbox{0.9cm}{\centering \small num.\\ classes} & \parbox{1.2cm}{\centering \small size trai-\\ ning set} & \parbox{0.97cm}{\centering\small size\\ test set} \\
      \midrule
      MNIST & $28 \times 28$ & 10 & 60000 & 10000\\
      MNIST-back-image & $28 \times 28$ & 10 & 12000 & 50000\\
      \LaTeX & $32 \times 32$ & 293 & 14650 & 9376\\
      \bottomrule
    \end{tabular}
  \end{table}

  \paragraph{Preprocessing} Every sample image was reshaped as a $l^2$-dimensional vector, and the average gray level of each component was subtracted from the data. The dimensionality of the resulting image vector was then reduced to 256 using PCA. To guarantee that all the selected components contributed uniformly to the patterns, we applied a random rotation to the principal subspace (see, e.g., \cite{Raiko2012}). We denote by $N_{\text{in}}=256$ the dimension of that subspace.

\subsection*{The architecture of the network and the training algorithm}

We map the preprocessed $N_{\text{in}}$-dimensional vector image, $\vec{s}$, into a higher dimensional space through the transformation
\[x_i = f(\vec{w}_i \cdot \vec{s}), \quad i=1,\ldots,N,\]
where $\vec{w}_{i}$ is an $N_{\text{in}}$-dimensional sparse random vector and $f(\cdot)$ is a nonlinear function. This is the transformation induced by a neural network with $N_{\text{in}}$ input units and $N$ output units with activation function $f(\cdot)$. More succinctly,
  \begin{equation}
    \vec{x} = f(\vec{W}^{T}\vec{s}),
    \label{eq:xrcn}
  \end{equation}
where $\vec{W}$ is a weight matrix of dimensions $N_{\text{in}} \times N$ formed by adjoining all the column weight vectors $\vec{w}_i$, and where $f(\cdot)$ acts componentwise, i.e., $f(\vec{x}) \equiv (f(x_1),\ldots,f(x_{N_{\text{in}}}))^{T}$.
The output of the random nonlinear transformation, $\vec{x}$, is used as the input to a linear $N_{C}$-class discriminant, consisting of $N_{C}$ linear functions of the type $y_{j} = \sum_{k=1}^{N} J_{jk} x_k$, with $j=1,\ldots, C$. More compactly,
  \begin{equation}
    \vec{y} = \vec{J} \vec{x},
    \label{eq:readout}
  \end{equation}
  where $\vec{y} = (y_1,\dotsc,y_{C})^{T}$, $\vec{J}$ is a $C \times N$ matrix, and $\vec{x}$ is given by Eq.~\eqref{eq:xrcn}. A pattern $\vec{x}$ is assigned to class $C_j$ if $y_j(\vec{x}) > y_k(\vec{x})$ for all $j \neq k$. The elements of $\vec{J}$ are learned offline by imposing a 1-of-$N_{C}$ coding scheme on the output: if the target class is $j$ then the target output $\vec{t}$ is a vector of length $N_{C}$ where all components are zero except component $t_j$, which is 1. For the offline training of weights we use the pseudoinverse, which minimizes the mean squared error of the outputs. This technique has been shown to be a good replacement for empirical minimization problems when the dataset is embedded in a random high-dimensional space, which is our case~\cite{huang2006,rahimi2008,tapson2013,le2013}.

  \section*{Neuromorphic chip implementation}
  The chip is composed of multiple identical cores, each of which consists of a neuromorphic circuit that comprises $n=256$ axons, $n$ neurons, and $n^2$ adjustable synapses~(\cite{merolla_etal_ieee2011,arthur_etal_ijcnn2012,merolla_etal_science2014}, see also Fig.~\ref{fig:basic}c). Each axon provides the inputs by feeding the spiking activity of one given neuron that may or not reside in the core. The incoming spiking activity to all $n$ axons in a core is represented by a vector of activity bits $(A_1(t),\dotsc,A_n(t))$ whose elements indicate whether or not the neurons associated with the incoming axons emitted a spike in the previous time step. The intersection of the the $n$ axons with the $n$ neurons forms a matrix of programmable synapses. The weight of active synapses is determined by the type of axon and the type of neuron the synapse lies on. Specifically, each core can contain up to four different types of axon, labeled $G_{j}=\{1,2,3,4\}$, whereas it can accommodate an unlimited number of neuron types, each of which having four associated synaptic weights $S_i = (S_i^{1}\!, \dotsc, S_i^{4})$. The strength of an active synapse connecting axon $j$ with neuron $i$ is $S_i^{G_j}$, that is, the axon type determines which weight to pick among the weights associated with neuron $i$. The net input received by neuron $i$ at time step $t$ is therefore $h_i(t) = \sum_{j=1}^{n} S_i^{G_j} B_{ij} A_j(t)$,
where $B_{ij}$ is 1 or 0 depending on whether the synapse between axon $j$ and dendrite $i$ is active or inactive.

At each time step the membrane potential $V_i(t)$ of neuron $i$ receiving input $h(t)$ is updated according to $V_i(t + 1) = V_i(t) - \beta + h_i(t)$, where $\beta$ is a constant leak. If $V_i(t)$ becomes negative after an update, it is clipped to 0. Conversely, when $V_i(t)$ reaches the threshold $V_{\text{thr}}$, the potential is reset to $V_{\text{reset}}$ and the neuron emits a spike, which is sent through the neuron's axon to the target core and neuron. This design implies that each neuron can connect to at most $n$ neurons, which are necessarily in the same core. The initial voltage of each neuron was initialized by drawing randomly and with equal probability from a set of 4 evenly spaced values from $V_{\text{reset}}$ to $V_{\text{thr}}$.

\paragraph{Signal-to-rate transduction}
The input to the neuromorphic chip consists of a set of spike trains fed to the neurons of the input layer. To transform the vector signal $\vec{s}$ into spike trains, we first shifted the signal by $\bar{s} = 3\sigma$, where $\sigma$ is the standard deviation across all signal components of all patterns. The shifted signal was then scaled by a factor $\nu_{\text{sc}}$ chosen to ensure moderate output rates in the RCN layer, and the result was linear-rectified to positive values. In short, the input rate $\nu_i$ associated with signal $s_i$ is $\nu_i = \nu_{\text{sc}} [ s_i + \bar{s}]_{+}$, $i=1,\dotsc,N_{\text{in}}$, where $[x]_{+}$ is $x$ if $x > 0$, or $0$ otherwise. The values $\nu_i$ were then used to generate regular spike trains with fixed inter-spike-interval $1/\nu_i$.

\paragraph{Basic architecture}
The circuit is divided in two functional groups, or layers, each of which comprises several cores. The first functional group is the RCN layer, which computes the random nonlinear expansion in Eq.~\eqref{eq:xrcn}. The second functional group computes the $C$-class discriminant $\vec{y} = \vec{J}\vec{x}$. The output of the classifier is just $\operatorname{argmax}_{j} y_j$, where $j$ runs over the $C$ possible categories.  The argmax operation was not computed by the chip, but was determined off-line by comparing the accumulated spike counts across all outputs. In the following, we describe the implementation of the two layers in more detail.

\paragraph{RCN layer}
We first set the dimensionality of the input to the number of available axons per core, i.e., $N_{\text{in}} = n = 256$. A convenient choice for $\mathbf{W}$ is a $n \times N$ matrix where each column is vector of zeros except for exactly $m < n$ nonzero entries, which are randomly placed and take a fixed integer value $w$. We took $m=26$, which corresponds to a connectivity level of around $0.1$. Lowering the connectivity has the advantage of decreasing energy costs by reducing the number of total spikes and active synapses, without impacting the classification performance. The random expansion was mapped in the chip by splitting the matrix $\mathbf{W}^{T}$ into $\lceil N / n\rceil$ submatrices of size $n \times n$, and using each submatrix as the (boolean) connectivity matrix $B_{ij}$ of a core.

 With this arrangement, each of the $N$ neurons distributed among the $\lceil N / n\rceil$ cores receives a sparse and random linear combination of signals. Specifically, the average current received by each RCN is
  \[h_i = \sum_{j=1}^{n} W_{ji} \nu_j, \quad i=1,\dotsc,N.\]
  % or, more compactly, $\mathbf{h} = \mathbf{W}^{T}\!\boldsymbol{\nu}$.

  A zero-th order approximation of the firing rate of a general \textsc{vlsi} neuron receiving a current $h_i$ is
  \begin{equation}
    r_i = \frac{[h_i - \beta]_{+}}{V_{\text{thr}} - V_{\text{reset}}},
    \label{eq:transferf}
\end{equation}
where $V_{\text{thr}}$ is the threshold for spike emission and $V_{\text{reset}}$ is the reset potential~\cite{fusi_mattia_neco1999}.

  We chose the parameters $w$ and $\beta$ to meet two criteria. First, we required the fraction of RCN{s} showing any firing activity (i.e., the coding level $f$) to be around $0.25$. This coding level is a good compromise between the need for discrimination and generalization, and it keeps finite-size effects at bay~\cite{barak_etal_jns2013}. Second, we required the distribution of activities across active RCN{s} to be sufficiently wide. Otherwise the information carried by the spiking activity of the RCN{s} is too imprecise to discriminate among patterns.

All the cores in the RCN layer receive exactly the same $n$-dimensional input signal.

\subsubsection*{Readout}
The readout matrix $\mathbf{J}$ was trained offline and mapped to the chip architecture as follows.

\paragraph{Weight quantization} Because the chip can hold only integer-valued synapses, we need to map the set of all components of $\vec{J}$ into an appropriate finite set of integers. We started clipping the synaptic weights within the bounds $(-4 \sigma, 4 \sigma)$, where  $\sigma$ is the standard deviation of the sample composed of all the components of $\mathbf{J}$. We then rescaled the weights to a convenient magnitude $J_{\text{max}}=28$ (see below), and rounded the weight values to the nearest integer. 

\begin{figure*}
    \centering
    \includegraphics{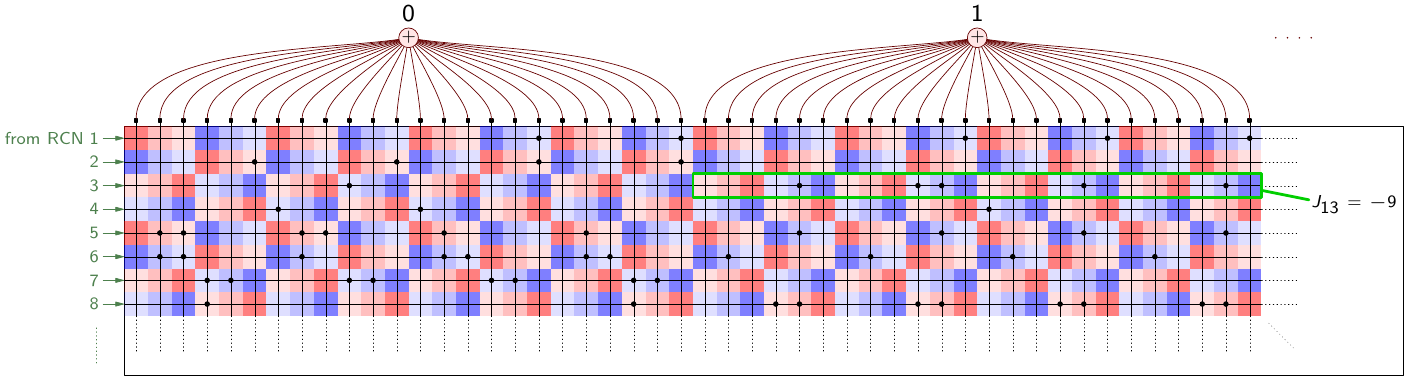}
    \caption{\textbf{Implementation  of the readout matrix in a core}. The diagram represents the first 8 input lines and first 48 dendrites (two output units) of a typical readout core. Under each axon-dendrite contact is a square that indicates the potential synaptic strength at the site: color indicates whether the connection is excitatory (red) or inhibitory (blue), while the saturation level represents the absolute value of the synaptic strength, which can be 1, 2, or 4 (low, medium, and high saturation, respectively). Only the sites marked with a dot are active. The green frame highlights all the synaptic contacts allocated for an arbitrary weight of the readout matrix, in this case $J_{13} = -9$, which is decomposed as the 4-term sum $-9 = -2 -3 -2 -2 = -010_2 - 011_2 - 010_2 - 010_2$. Note that in this particular axon the ordering of weights is $2^{0}, 2^{1}, 2^{2}$ (rightmost bit is the most significant).
    \label{fig:readout_arch}}
  \end{figure*}

\paragraph{Weight assignment}
The TrueNorth connectivity constraints dictate that each RCN can project to only one axon, meaning that there are at most $n=256$ synaptic contacts available to encode the $C=10$ weights, $J_{0i},\dotsc,J_{9i}$ associated with the $i$-th RCN. We allocated 24 contacts per class and per axon (see Fig.~\ref{fig:readout_arch}). Each of these 24 contacts were divided in four groups comprising 6 weights each, with values $1, 2, 4, -1, -2, -4$. This allowed us to represent any integer weight from $-28$ to $28$ (each of the 4 groups encodes a maximum weight of $7$, sign aside). To distribute any weight value $w$ across the available synaptic contacts, we decomposed $w$ in a sum of four terms, given by the integer division of $w$ by 4 with the remainder spread evenly across terms (Ex: $19 = 4 + 5 + 5 + 5$). Each of such values was assigned to one group, represented in base 2, and mapped to a pattern of active-inactive synapses according to the weight associated with each axon-dendrite intersection. Positive and negative weights, as well as strong and week weights, were balanced along a dendrite by changing the sign and order of the weights in the crossbar (see alternating colors and saturations in Fig.~\ref{fig:readout_arch}).

\paragraph{Negative threshold} For the readout to work properly, the firing activity of readout neurons must be proportional to the linear sum of the inputs from the RCNs. This requires neurons to operate in the linear regime of their dynamic range, a regime that can be enforced by lowering the threshold $\beta_{\text{out}}$ of readout neurons. We set $\beta_{\text{out}}<0$, which is equivalent to adding a constant positive current to each neuron. If the current-to-rate transduction function were the threshold-linear function of Eq.~\eqref{eq:transferf}, the baseline activity induced by this constant current would be $|\beta_{\text{out}}|/(V_{\text{thr}} - V_{\text{reset}})$ per readout neuron. The contribution of this background signal should be subtracted from the readout outputs if one wants to get the equivalent to Eq.~\eqref{eq:readout}, although the step is unnecessary if one only wishes to compare output magnitudes (as we implicitly do in order to find the maximal output).

\subsection{Support Vector Machines}
We trained SVMs to perform multiclass classifications based on a one-vs-all scheme, so that the number of output units coincides with the number of classes (as in the neural classifier).
% Details. Kernel. Optimization procedures for minimizing the number of support vectors.
SVMs were evaluated using arc-cosine kernels, whic mimic the computation of large feedforward networks with one or more layers of hidden nonlinear units~\cite{cho2010}. For our particular architecture, based on one hidden layer built with threshold-linear units, the kernel is $k(\vec{x},\vec{y}) = \|\vec{x}\| \|\vec{y}\| J_1(\theta)$, where $J_{1}(\theta) = \sin \theta + (\pi - \theta) \cos \theta$ and $\theta$ is the angle between the inputs $\vec{x}$ and $\vec{y}$.

We considered three types of SVM. For the standard SVM we used the open library \texttt{libsvm}~\cite{chang_lin_acm2011}, which we patched to include the arccos kernel. The other two SVMs reduce the number of support vectors without sacrificing performance substantially. One of such algorithms is \texttt{primalSVC}, which selects greedily the basis functions by optimizing the primal objective function~\cite{keerthi_etal_jmlr2006}. The other method is based on the so-called Cutting-Plane Subspace Pursuit algorithm, which reduces the number of support vectors by using basis functions that, unlike standard SVMs, are not necessarily training vectors~\cite{joachims_yu_ml2009}. Such method is implemented in the library \texttt{SVMperf}. Unlike the other two classifiers, \texttt{SVMperf} used RBF kernels instead of arccos kernels.

\subsection*{Estimation of the IBM chip energy consumption}
\label{sec:ibmchipenergy}
% The total energy consumption is
% \[ E = (P_{\text{base}} + P_{\text{read}} n_{\text{syn}} ) \Delta t + n_{\text{sp}} E_{\text{sp}} + n_{\text{upd}}\dots\]
  % ORE(FAVE,NSYN)         = PBASE +ENRN ×FAVE +PREAD ×NSYN +ESYN ×FAVE ×NSYN
                        % = 15.9μW +28nJ ×FAVE +2.5nW ×NSYN +0.3nJ ×FAVE ×NSYN

                        % where
                        % • PBASE = 15.9μW is the core’s baseline power (passive plus active)
                        % • EN RN = 28nJ is energy for 256 neuron spikes (one for each neuron on a core)
                        % • PREAD = 2.5nW is the power for reading 256 active synapses per time step
                        % (one for each neuron)
                        % • ESY N = 0.3nJ is the energy for 256 synaptic updates (one for each neuron)
                        % • FAV E is the average spike rate (Hz) of all 256 neurons in a core (valid
                        % from 0 to 200Hz)
                        % • NSY N is the average number of active synapses per neuron in a core (1 to
                        % 256)

                        % All power numbers assume a time step of 1ms.
The energy consumption of the IBM chip was estimated from the TrueNorth~specifications~\cite{merolla_etal_science2014}. The total energy consumption comprises the baseline energy ($15.9\,\mu\text{W}$ per core), the energy to emit spikes ($109\uni{pJ}$ per spike), the energy needed to read active synapses ($10.7\uni{pJ}$ per active synapse), and the energy necessary to update membrane potentials ($1.2\uni{pJ}$ per neuron). We ignored the input-output energy needed to transmit spikes off chip and receive spikes on chip. These numbers provide a reasonable estimate of the energy consumption of systems with a conservative supply voltage of $0.775\uni{V}$; most chips operate near or below this estimate. For a setup with $2^{14}$ RCNs, 26 dendrites per class, and 10 classes, the power was about $2.08\uni{mW}$, $95\%$ of which corresponds to the baseline power.

\paragraph{Scaling of the energy with the number of classes} The estimation was based on the energy cost of the simulated classifications of the MNIST dataset, and extrapolated to the designs required by an increasing number of classes. As the number of classes $C$ increases, so does the number of readout neurons necessary to perform a classification and, therefore, so does the required number of readout cores. Specifically, if we assign $s_c$ synaptic contacts per axon and per class, we will need a total of $s_c C$ output lines. These output lines need to be connected to all the $N$ neurons through the input lines of the readout cores . Because each readout core can accomodate 256 output lines, connected to 256 input lines, the total number of readout cores will be $\lceil N / 256 \rceil \lceil s_c C / 256 \rceil$ ($\lceil \cdot \rceil$ indicates the ceiling function). In principle the number of RCN cores will be simply $\lceil N / 256 \rceil$. However, each RCN should project to $\lceil s_c C / 256 \rceil$ cores, which implies that each RCN core must be cloned $\lceil s_c C / 256 \rceil$ times due to the fan-out constraint---each RCN can project to only one core. The total number of cores is therefore $N_{\text{cores}} = 2 \lceil N / 256 \rceil \lceil s_c C / 256 \rceil$, where the factor 2 accounts for the contributions of both the readout and the RCN cores. 

The total number of spikes emitted was estimated from the reference value we got from the chip simulation (for 10 classes, $N = 2^{14}$, $s_c=24$, and $500\uni{ms}$ of classification time), scaled appropriately for the new  $N_{\text{cores}}$. More concretely, if we denote by $n_{\text{sp}}^{0}$ the number of spikes emitted during our reference simulation, the number of emitted spikes in a general case is $n_{\text{sp}} = n_{\text{sp}}^{0} \lceil s_c C / 256 \rceil (T / 500) (N / 2^{14})$, where $T$ is the duration of the simulation in milliseconds. We chose this duration to be $T=108\uni{ms}$, which is the average classification time of the chip implementation the MNIST dataset, when the spike difference is 80 spikes and which yields only 0.1\% less in performance than in the fixed-duration case (97.2\% vs 97.3\%). With $T$ and the estimated values of $N_{\text{cores}}$ and $n_{\text{sp}}$, it is straightforward to compute the energy consumption according to the values given in the previous paragraph.

\section{Energy consumption in von Neumann digital machines} 

\label{sec:digital_machines}
\paragraph{Configuration} The runtime and power of microprocessors with von Neumann architectures were estimated with the recently developed simulators GEM5 (gem5.opt 2.0)~\cite{binkert_etal_can2011}
and McPAT (ver.\ 1.2)~\cite{li_etal_microarch2009}. For the estimation we used an architecture configuration similar to that of the recent Intel Core\textsuperscript{TM} i7 processors \cite{intel_i7}, which incorporate state-of-the-art CMOS technology. Specifically, we used an x86\_64, O3, single core architecture at 2.66 GHz clock frequency, with 32KB 8-way L1-i and 32KB 8-way L1-d caches, 256KB 8-way L2 cache, 64B cache line size, and 8GB DDR3 1600 DRAM. Channel length was $22\uni{nm}$, HP type, using long channel if appropriate. VDD was 0.9V, so slightly higher than the 0.775 V used for the IBM chip. However, could we use the same voltage in Intel i7 simulator, the energy consumption would be lower by a factor $(0.775/0.9)^2=0.74$. This 26\% reduction would not change the main conclusions about the energy consumption gap between the IBM chip and the conventional von Neumann digital machines, which is 2--3 orders of magnitude.

\begin{figure}
  \centering
  \includegraphics{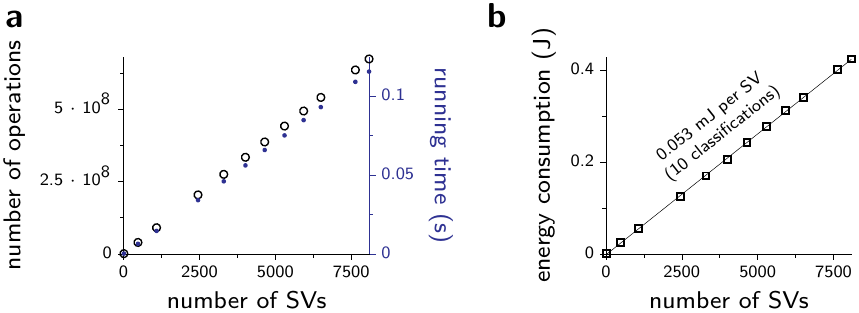}
  \caption{\textbf{Simulation of a digital support vector machine.} \textbf{a} Number of operations (black circles, left ordinate) and runtime (blue dots, right ordinate) required by a digital SVM to classify 10 test patterns from the MNIST dataset, as a function of the number of support vectors. The SVM performance was estimated with a simulator of the Intel i7 processor. \textbf{b} Energy consumption associated to the datapoints shown in \textbf{a} (squares). The straight line is a least-square fit.
  \label{fig:mingoo_data}}
\end{figure}

\paragraph{Simulations} The benchmark was the test phase of the SVMs, already trained. Simulations showed that a modern microprocessor based on a von-Neumann architecture takes $115.5\uni{ms}$ to evaluate the test set with 8087 SVs, while consuming $424.6\uni{mJ}$ (DRAM energy consumption not included). When we varied the number of support vectors from 9 to 8087, both the runtime and energy consumption grew proportionally to the number of SVs, while the power was roughly constant due to the fixed hardware configuration (see Fig~\ref{fig:mingoo_data}). 
To estimate how the energy used by von Neumann digital SVMs scales with the number of classes, we ran another set of simulations with Intel i7 simulator, this time varying both the number of support vectors and the number of classes in the classification problem. This step was necessary to determine the overhead incurred when we increase the number of output units. For a given number of classes, the energy cost per support vector was estimated from the least-square fit of the energies against the number of support vectors.

% {\bfseries Should we include the plots by Mingoo here?}
% \begin{tabular}{r r . r r}
  % \toprule
  % \parbox{1.2cm}{\centering number\\ SVs} & \parbox{1.2cm}{\centering running\\ time (s)} &
  % \multicolumn{1}{c}{\parbox{1.8cm}{\centering number operations\\ (millions)}} &
% \parbox{1.0cm}{\centering power\\ (W)} & \parbox{1.1cm}{\centering
% energy\\ (J)}\\
% \midrule
   % 9 & 0.00017 & 0.79 & 3.243 & 0.00054 \\
 % 473 & 0.00638 & 39.5 & 3.774 & 0.02408 \\  
% 1086 & 0.01454 & 90.5 & 3.794 & 0.05516 \\ 
% 2451 & 0.03406 & 204  & 3.690 & 0.12568 \\ 
% 3300 & 0.04601 & 275  & 3.690 & 0.16978 \\ 
% 4014 & 0.05611 & 334  & 3.670 & 0.20592 \\ 
% 4649 & 0.06587 & 387  & 3.678 & 0.24227 \\ 
% 5302 & 0.07513 & 442  & 3.687 & 0.27700 \\ 
% 5932 & 0.08482 & 494  & 3.673 & 0.31154 \\ 
% 6499 & 0.09308 & 541  & 3.663 & 0.34095 \\ 
% 7637 & 0.10909 & 636  & 3.674 & 0.40080 \\ 
% 8087 & 0.11560 & 674  & 3.673 & 0.42460 \\ 
% \bottomrule
% \end{tabular}

\paragraph{Mobile processor} We also investigated the runtime and energy consumption of a more energy-efficient but slower mobile microprocessor performing the same target workload. The architecture configuration was: ARMv7, O3, single core, 1GHz CPU clock frequency, 32kB 4-way L1i and 32kB 4-way L1d caches, and 128kB 8-way L2 cache, which is similar to the architecture of ARM Cortex-A9~\cite{armcortex}. The technology node ($22\uni{nm}$) and simulators were the same as in the experiment with the microprocessor mimicking Intel Core i7. For the benchmark code with the largest number of SVs, the task required $1.2\cdot10^{10}$ operations that 
took $6.35\uni{s}$ at a cost of $7.34\uni{J}$.

\paragraph{Discussion on Intel Xeon Phi} Massively parallel architectures have gained a significant amount of attention to improve the throughput and power efficiency of the high-performance computing (HPC) technology, in response to the relatively stagnated improvement in clock frequency. The Xeon Phi coprocessor, recently developed by Intel, is one of such efforts~\cite{chrysos_hc2012}. It integrates more than 50 CPU cores together with L1/L2 caches, network-on-chips, GDDR memory controller, and PCIe interface. Each core supports up-to 4-thread in-order operation and the 512b SIMD VPU (Vector processing unit). While the runtime and energy-consumption of the coprocessor are highly dependent on the target workloads, several recent investigations quantified the performance and energy-efficiency. In the high-performance configuration, the system integrating Xeon and Xeon Phi shows the throughput of 100 Tera floating-point operations (flop) per second, the power consumption of $72.9\uni{kW}$, marking the energy efficiency of $0.74\uni{nJ}/\text{flop}$~\cite{chrysos_hc2012}. The classification benchmark codes (with the largest number of SVs) require 0.02235 Gigaflop on the desktop processor configuration similar to Intel Core i7. At a first order approximation, therefore, the Xeon and Xeon Phi-based system takes $0.2235\,\mu\mathrm{s}$ and uses $16.5\uni{mJ}$ per classification. This energy consumption seems significantly lower than the one of the Intel Core i7, and very close to its lower bound, which is approximately $3\uni{mJ}$. However, one should keep in mind that the energy is grossly underestimated, as not only we ignored the energy needed for the RAM, but we also neglected the cost of the non floating point operations, which are approximately twice as many as the floating point operations. For all these reasons it is difficult to compare the energy consumption for the Xean Phi to the Intel Core i7. In any case, even for our very conservative energy consumption estimate, the IBM chip remains significantly more energy efficient.

\bibliography{ms}
\end{article}
\end{document}